\documentclass[12pt,english]{article}
\usepackage{times}
\usepackage[T1]{fontenc}
\usepackage[latin1]{inputenc}
\usepackage{geometry}
\usepackage{array}
\usepackage{graphicx}
\usepackage{setspace}
\makeatletter

\providecommand{\tabularnewline}{\\}

\newcommand{\lyxaddress}[1]{
\par {\raggedright #1
\vspace{1.4em}
\noindent\par}
}

\usepackage{babel}
\makeatother
\begin{document}

\title{Functional characteristics of a double positive feedback loop coupled
with autorepression}

\author{Subhasis Banerjee and Indrani Bose}

\maketitle

\lyxaddress{\begin{center}
Department of Physics, Bose Institute, 93/1,  A. P. C Road, Kolkata-700009,
India
\par\end{center}}

\begin{abstract}
We study the functional characteristics of a two-gene motif consisting
of a double positive feedback loop and an autoregulatory negative
feedback loop. The motif appears in the gene regulatory network controlling
the functional activity of pancreatic $\beta$-cells. The model exhibits
bistability and hysteresis in appropriate parameter regions. The two
stable steady states correspond to low (OFF state) and high (ON state)
protein levels respectively. Using a deterministic approach, we show
that the region of bistability increases in extent when the copy number
of one of the genes is reduced from two to one. The negative feedback
loop has the effect of reducing the size of the bistable region. Loss
of a gene copy, brought about by mutations, hampers the normal functioning
of the $\beta$-cells giving rise to the genetic disorder, maturity-onset
diabetes of the young (MODY). The diabetic phenotype makes its appearance
when a sizable fraction of the $\beta$-cells is in the OFF state.
Using stochastic simulation techniques, we show that, on reduction
of the gene copy number, there is a transition from the monostable
ON to the ON state in the bistable region of the parameter space.
Fluctuations in the protein levels, arising due to the stochastic
nature of gene expression, can give rise to transitions between the
ON and OFF states. We show that as the strength of autorepression
increases, the ON$\rightarrow$OFF state transitions become less probable
whereas the reverse transitions are more probable. The implications
of the results in the context of the occurrence of MODY are pointed
out..

P.A.C.S. Nos.: 87.18Cf, 87.18Tt, 87.18Vf

\newpage
\end{abstract}

\section{Introduction}

Positive and negative feedback loops are frequently-occurring motifs
in gene transcription regulatory networks and signaling pathways \cite{key-1,key-2}.
The components of a feedback loop are genes, proteins and other molecules
which are connected by regulatory interactions. Depending on the components
and their interactions, feedback loops have distinct roles in diverse
regulatory systems. A regulatory interaction is positive (negative)
if an increase in the amount or activity of one component increases
(decreases) the amount or activity of its interaction partner. A feedback
loop is positive (negative) if the number of repressing interactions
is zero or even (odd). A large number of experiments and theoretical
studies elucidate the major functional characteristics of feedback
loops with simple structure \cite{key-1,key-2,key-3,key-4,key-5,key-6,key-7,key-8}.
Positive feedback in a gene transcription regulatory network (GTRN)
tends to enhance protein levels whereas negative feedback favours
homeostasis ,i.e., maintenance of proteins at a desired level. The
simplest feedback loop has only one component which is thus self-regulating.
For such a motif in a GTRN, a protein promotes / represses its own
production via autoactivation / autorepression of the expression of
its gene. A positive feedback loop with two components and two regulatory
interactions is of two types: double negative and double positive.
Again, considering a GTRN, the protein products of the two genes in
a double negative feedback loop repress each other's synthesis. The
construction of a synthetic circuit, the genetic toggle, is based
on this motif \cite{key-9}. The double positive feedback loop is
defined by two genes, the protein products of which promote each other's
synthesis. There are several examples of two-component positive feedback
loops in natural cellular networks \cite{key-1,key-2}, a prominent
example being the cell division cycle, the regulatory network of which
contains both double positive and double negative feedback loops \cite{key-10}.
In this case, the loops control enzymatic activity. The double negative
feedback loop, because of its more common occurrence, has been extensively
studied in contrast to the double positive feedback loop. 

The next stage of complexity in feedback loops involves linked positive
and negative feedback loops \cite{key-2,key-11,key-12,key-13}. The
key variables in the dynamics of feedback loops are the concentrations
of the component molecules. In the case of a GTRN, these may be the
protein concentrations. In a deterministic description, the time evolution
of the concentrations is determined by solving a set of coupled differential
equations, the number of equations being equal to the number of variables.
In reality, the biochemical events associated with gene expression
are probabilistic in nature and this is reflected in the presence
of fluctuations (termed noise) around mean protein levels \cite{key-14}.
A stochastic description of time evolution is thus more appropriate.
A single positive feedback loop has a tendency to amplify noise, also
the time taken to reach the steady state protein level is longer than
that in the case of an unregulated gene \cite{key-1,key-2}. Interlinking
of two positive feedback loops with slow and fast dynamics results
in a switch with rapid activation and slow deactivation times and
a marked resistance to noise in the upstream signaling pathway \cite{key-11}.
Addition of a single negative feedback loop leads to rapid deactivation
in the absence of the signal which activates the switch \cite{key-12}.
The combination of positive and negative feedback loops may give rise
to excitability with transient activation of protein levels. Recent
experiments suggest that competence development in B. subtilis is
achieved via excitability \cite{key-15}.

\begin{figure}
\begin{centering}
\includegraphics[scale=0.6,bb = 0 0 200 100, draft, type=eps]{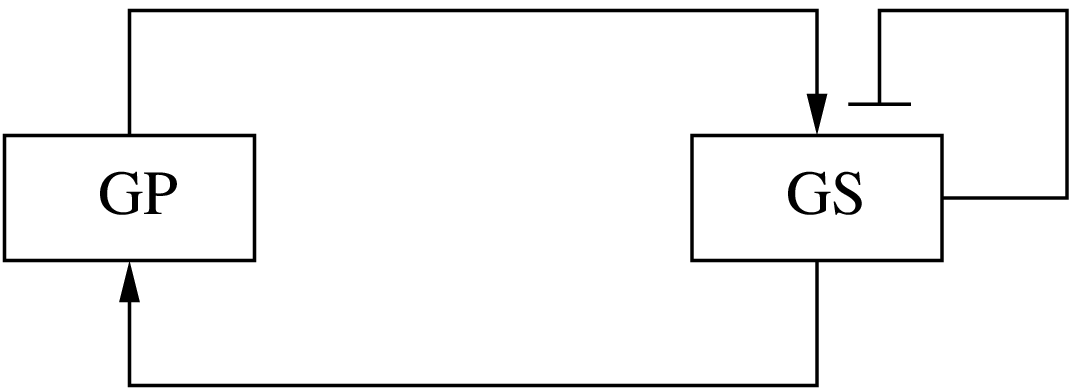}
\par\end{centering}

\caption{The two-gene network model. The protein products of the $GP$ and
$GS$ genes activate each other's synthesis. There is also an autorepressive
loop in which the proteins of the $GS$ gene repress their own synthesis.}
\end{figure}

In this paper, we study the functional characteristics of a two-gene
double positive feedback loop coupled with autorepression of the expression
of one of the genes. The major motivation for studying this specific
motif is its presence in the GTRN controlling the pancreatic $\beta$-cell
function \cite{key-16}. The hormone insulin is a small protein that
is synthesized in the $\beta$-cells and secreted when an increase
in the blood glucose level is sensed. Glucose metabolism releases
energy needed by cells to do useful work. Insulin is necessary to
metabolize glucose and thereby control its level in the blood. Diabetes
occurs due to an excessive accumulation of glucose in the blood brought
about by an insufficient production of or reduced sensitivity to insulin.
The core of the $\beta$-cell transcriptional network consists of
a double positive feedback loop in which the transcription factors
$HNF-1\alpha$ and $HNF-4\alpha$, belonging to the nuclear hepatocyte
family, activate each other's synthesis. There is also some evidence
that $HNF-4\alpha$ autorepresses its own synthesis \cite{key-16}.
Mutations in the transcription factors $HNF-1\alpha$ and $HNF-4\alpha$
give rise to a type of diabetes known as maturity-onset diabetes of
the young (MODY) which has an early onset with age less than usually
$25$ years. There are six different forms of MODY of which MODY 1
and MODY 3 are caused by mutations in the genes $hnf-4\alpha$ and
$hnf-1\alpha$ respectively \cite{key-34}. The structure of the regulatory
network, of which the two genes $hnf-1\alpha$ and $hnf-4\alpha$
are integral components, is not fully known. A partial structure of
the complex network is shown in \cite{key-16,key-34} involving the
genes $hnf-1\alpha,$ $hnf-4\alpha,$ $shp,$ $hnf-1\beta,$ $hnf-3\beta,$
$hnf-3\gamma,$ $hnf-4\gamma,$ and $pdx-1.$ The genes collectively
control the transcription of a number of important genes involved
in glucose metabolism in the $\beta-$cell. These include the glucose
transporter $2$ ($Glut-2)$ gene, the glucokinase gene encoding the
glycolytic enzyme glucokinase which acts as glucose sensor and also
the insulin gene. Odom et al. \cite{key-35} combined chromatin immunoprecipitation
assays with promoter microarrays to gain insight on the regulatory
circuits formed by $hnf-1\alpha$ and $hnf-4\alpha.$ Both the proteins
are found to control the activity of a large number of target genes
in the $\beta-$cell. This recent finding as well as earlier experiments
\cite{key-16} indicate that the $hnf-1\alpha$ and $hnf-4\alpha$
genes play a prominent role in the pancreatic $\beta-$cell function.
Mutations in the genes give rise to MODY resulting in the impairment
of glucose-stimulated insulin secretion. Several experiments \cite{key-16}
provide clues on the possible molecular origins of MODY. The cross-regulatory
interactions between $HNF-1\alpha$ and $HNF-4\alpha$ are switched
on as pancreatic $\beta$-cells receive the signals to differentiate.
The double positive feedback loop has the potential for bistability,
i.e., two stable steady states. The two states are a basal state in
which the two genes have low activity and an activated state which
corresponds to high protein levels. The states are analogous to the
OFF and ON states of a switch. Normal functioning of the pancreatic
$\beta$-cells requires the two-gene feedback loop to be in the ON
state. The circuit operation is, however, vulnerable to decreased
gene dosage caused by mutations (in a diploid organism each gene has
two identical copies). Genetic disorders, termed haploinsufficiency,
are known to occur due to reduced gene dosage resulting in decreased
protein levels \cite{key-17,key-18,key-19,key-20}. Gene expression
noise increases the probability that a protein level falls below a
threshold value so that the protein amount is insufficient for meaningful
activity. The loss of vital protein functions is responsible for the
occurrence of genetic disorders. MODY, brought about by reduced gene
dosage, is thus an example of haploinsufficiency \cite{key-16}. We
construct a mathematical model to study the dynamics of the core circuit
consisting of a double positive feedback loop coupled with autorepression
of the $hnf-4\alpha$ gene. We use both deterministic and stochastic
approaches to identify the functional features of the motif and discuss
their possible relevance in the occurrence of MODY

\section{Deterministic Approach}

The circuit diagram of the motif to be studied is shown in the figure
1. $GP$ and $GS$ represent the genes $hnf-1\alpha$ and $hnf-4\alpha$
respectively. The arrow sign denotes activation by the appropriate
protein product and the hammerhead sign denotes repression. The chemical
kinetic schemes corresponding to the expression of genes $GP$ and
$GS$ are shown in figures 2(a) and 2(b). The protein products of
$GP$ and $GS$ are denoted by $P$ and $S$ . We assume that the
regulation of gene expression is mediated by the protein dimers $S2$
and $P2$, $K_{P}$ and $K_{S}$ being the binding constants of dimerization.
For each gene, there are two rates of protein synthesis: a basal rate
( rate constants $J_{P0}$ and $J_{S0}$) and an activated rate (
rate constants $J_{P}$ and $J_{S}$). In the second case, protein
synthesis occurs in the activated state of the gene ($GP^{*}$ and
$GS^{*}$) attained via the binding of protein dimers $S2$ and $P2$
to the genes $GP$ and $GS$ respectively. The associated binding
constants are $K_{PP}$ and $K_{SS}$ . The rate constants for protein
degradation are $\gamma_{P}$ and $\gamma_{s}$ with $\phi$ denoting
the degradation product. Dimer degradation is not taken into account
as its rate is few-fold lower than the degradation rate of protein
monomers. For the gene $GS$, there is an extra biochemical event
representing autorepression. The dimers $S2$ and $P2$ bind the promoter
region of the gene GS competitively, i.e., the binding of one type
of dimer excludes the binding of the other type. When the dimer $S2$
binds at $GS$, there is complete repression. The binding constant
is denoted by $K_{R}$. 

\begin{figure}
\begin{centering}
\includegraphics[clip,scale=0.5,bb = 0 0 200 100, draft, type=eps]{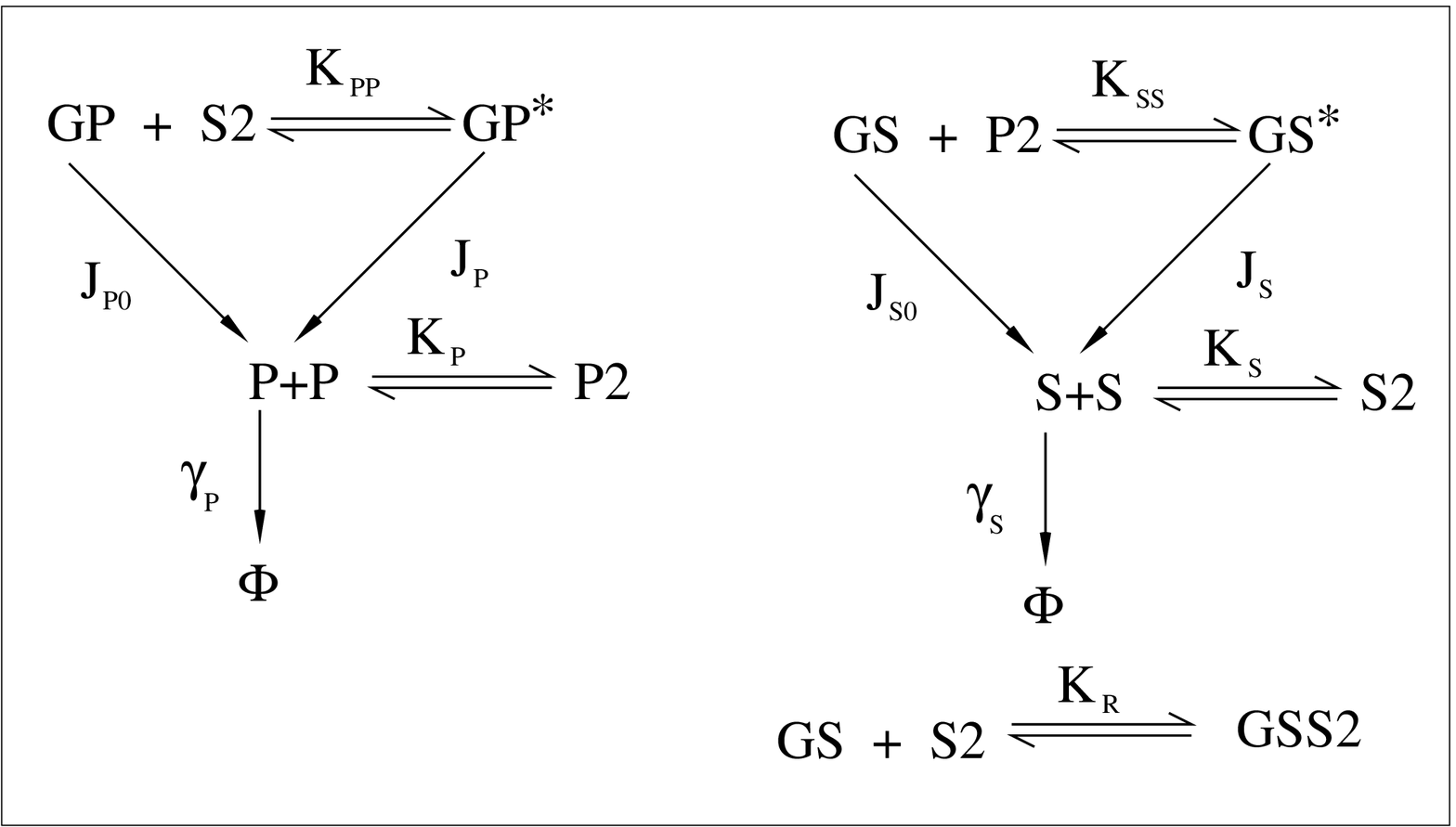}
\par\end{centering}

\caption{The reaction kinetic scheme of the two-gene model. The meanings of
the symbols are explained in the text.}
\end{figure}

The protein concentrations $S$ and $P$ are the dynamical variables
in the system. The time scale of binding events, in general, is much
faster than that of protein synthesis and degradation. The bound complexes
thus reach the steady state at an earlier time point. Taking this
into account, the differential rate equations describing the time
evolution of the protein concentrations $S$ and $P$ are: 

\begin{equation}
\frac{dS}{dt}=J_{S}\, GS^{*}+J_{S0}\, GS-\gamma_{S}\, S\end{equation}
\begin{equation}
\frac{dP}{dt}=J_{P}\, GP^{*}+J_{P0}\, GP-\gamma_{P}\, P\end{equation}
with \begin{equation}
GS^{*}=\frac{n_{S}\; T\,\left(\frac{P}{M}\right)^{2}}{1+T\,\left(\frac{P}{M}\right)^{2}},\;\; M^{2}=\frac{1}{K_{SS}\, K_{P}},\;\; T=\frac{1}{1+K_{R}\, K_{S}\, S^{2}}\end{equation}
\begin{equation}
GP^{*}=\frac{n_{P}\;\left(\frac{S}{N}\right)^{2}}{1+\left(\frac{S}{N}\right)^{2}},\;\; N^{2}=\frac{1}{K_{PP}\, K_{S}}\end{equation}

There are two conservation equations for the total concentrations
$n_{S}$ and $n_{P}$ of the genes $GS$ and $GP$. 

\begin{equation}
n_{S}=GS+GS^{*}+GSS2\end{equation}

\begin{equation}
n_{P}=GP+GP^{*}\end{equation}

After an appropriate change in variables \begin{equation}
u=\frac{S}{J_{S0}/\gamma_{S}},\; v=\frac{P}{J_{P0}/\gamma_{P}},\;\tau=\gamma_{S}\: t\end{equation}
the differential rate equations (1) and (2) are transformed into 

\begin{equation}
\frac{du}{d\tau}=n_{S}\frac{1+\eta\,\beta\, v^{2}}{(1+\mu\, u^{2})+\beta\, v^{2}}-u\end{equation}

\begin{equation}
\frac{dv}{d\tau}=n_{P}\frac{1+\xi\,\alpha\, u^{2}}{1+\alpha\, u^{2}}-v\end{equation}

The different parameters are given by \begin{equation}
\eta=\frac{J_{S}}{J_{S0}},\;\;\xi=\frac{J_{P}}{J_{P0}},\;\;\mu=\left(\frac{J_{S0}}{\gamma_{S}}\right)^{2}K_{S}\, K_{R},\;\;\alpha=\left(\frac{J_{S0}}{\gamma_{S}}\right)^{2}K_{PP}\, K_{S},\;\;\beta=\left(\frac{J_{P0}}{\gamma_{P}}\right)^{2}K_{SS}\, K_{P}\end{equation}

The variable $\tau$ is dimensionless whereas the variables $u$ and
$v$ have the dimensions of concentration expressed in units of {[}nm].
The parameters $\eta$ and $\xi$ are dimensionless while the parameters
$\mu$, $\alpha$ and $\beta$ are expressed in units of $\frac{1}{[nm]^{2}}$.
The dimensions of $n_{s}$ and $n_{p}$ are in units of $[nm]$ with
one gene copy corresponding to approximately $1[nm]$. From now on,
the units will not be explicitly mentioned. Table 1 displays all the
parameters and rescaled parameters of the two-gene model as well as
their meanings and defining formulae. 

\begin{figure}
\begin{centering}
\includegraphics[clip,scale=0.35,angle=-90,bb = 0 0 200 100, draft, type=eps]{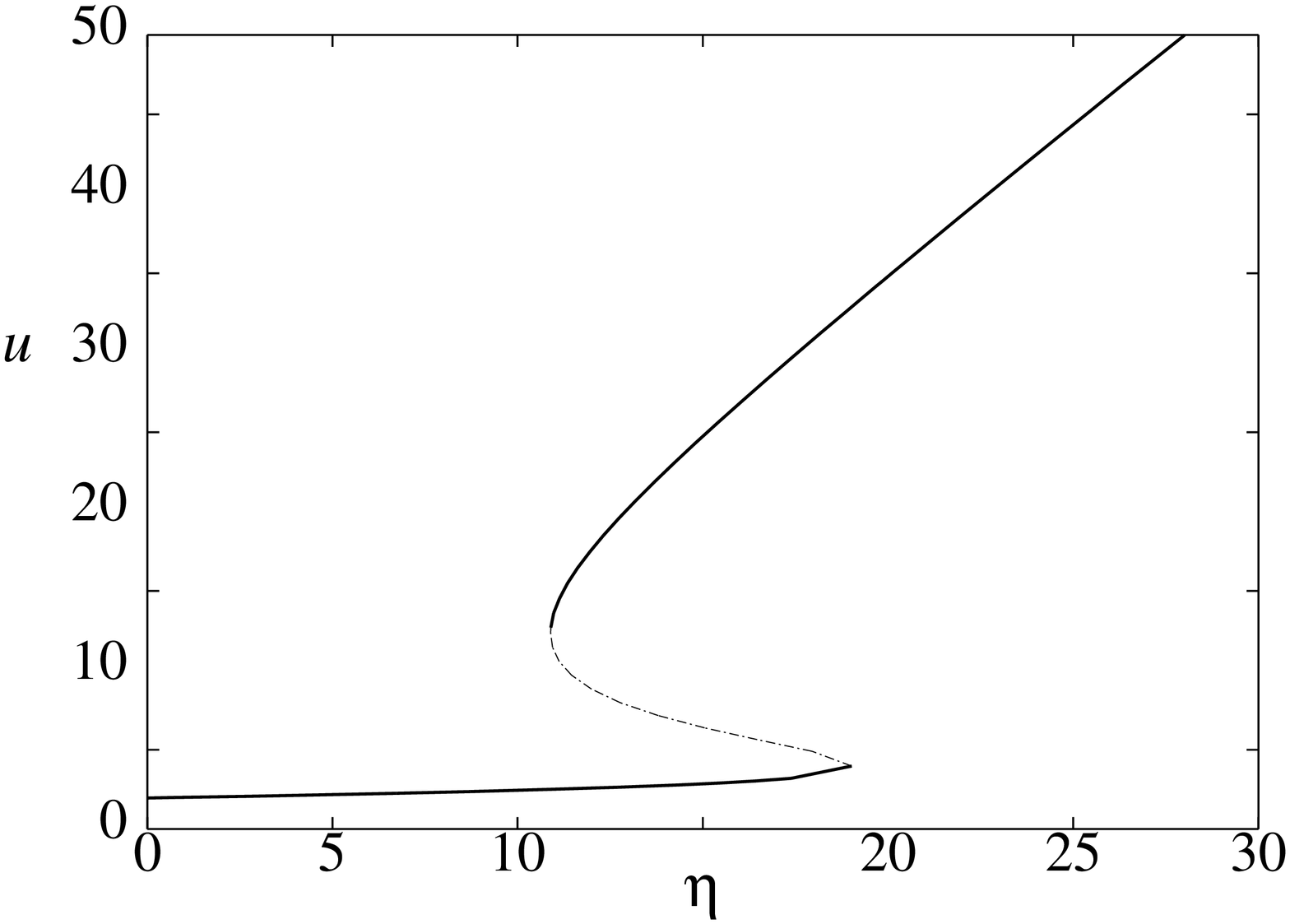}\includegraphics[clip,scale=0.35,angle=-90,bb = 0 0 200 100, draft, type=eps]{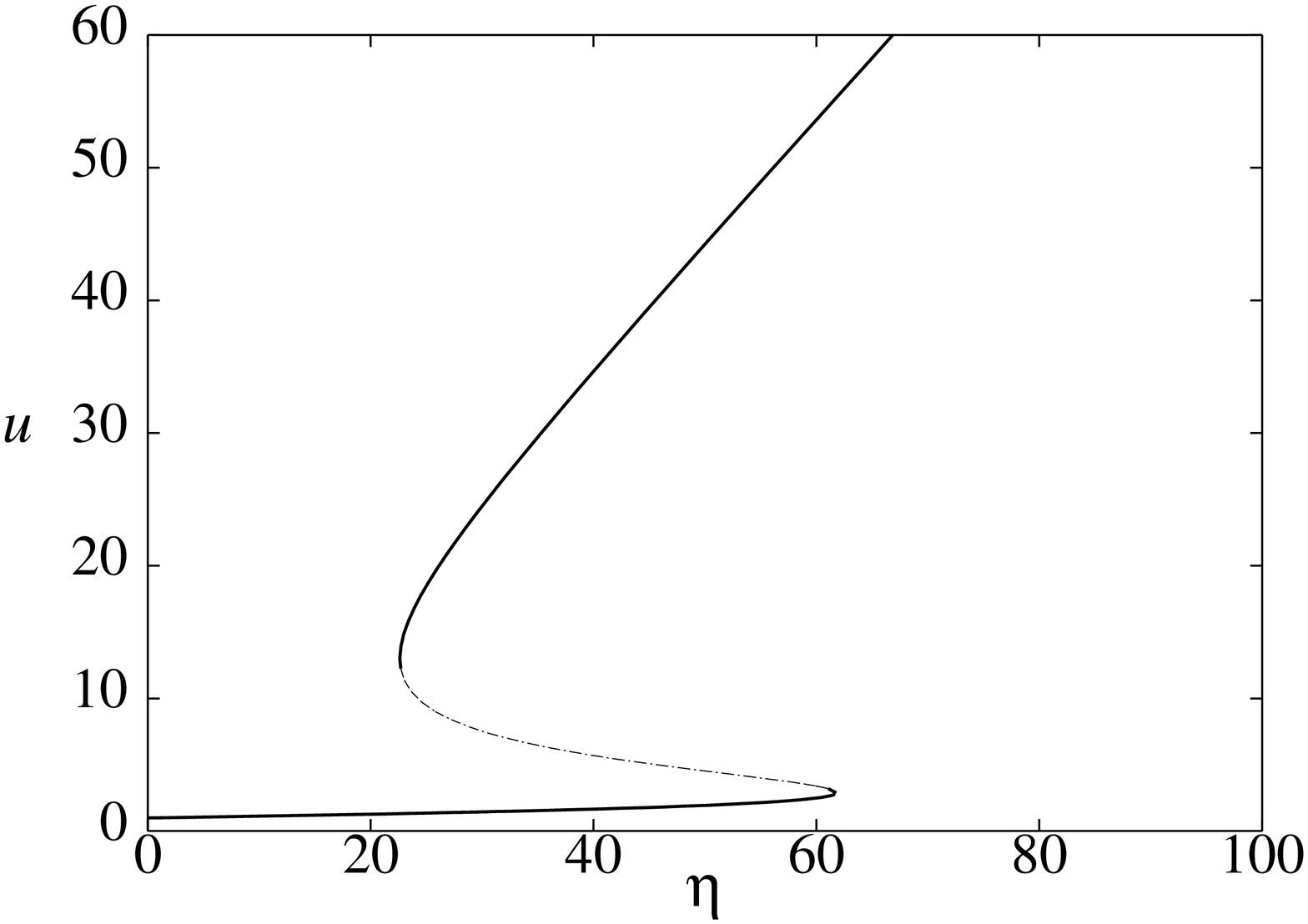}
\par\end{centering}

\caption{$u$ versus $\eta$ curves showing bistability and hysteresis. The
solid (dashed) lines represent stable (unstable) steady states for
gene copy numbers (a) $n_{P}=2$, $n_{S}=2$ and (b) $n_{P}=2$, $n_{S}=1$
. The parameter $\mu$, a measure of the autorepression strength,
is zero.}
\end{figure}

We use the software package XPPAUT  \cite{key-21}  to probe the dynamics
of the double positive feedback loop and the effect of autorepression
of the $S$ proteins on the dynamics . We focus on how the steady
state value of u (rescaled concentration of $S$ proteins) changes
as a function of the different parameters in equations (8) and (9).
In the steady state, the rates of change $\frac{du}{d\tau}$ and $\frac{dv}{d\tau}$
are zero. Figure 3(a) shows a plot of $u$ versus $\eta$ when the
autorepression strength given by $\mu$ is zero. The other parameters
have values $n_{S}$ = $n$$_{P}=2$, $\xi=30.0$, $J_{S0}=J_{P0}=2.0$,
$\gamma_{S}=\gamma_{P}=1.0$ and $\alpha=\beta=0.002857$. The plot
shows that a region of bistability separates two region of monostability.
The two stable states in the bistable region correspond to low and
and high values of $u$ . In this region and at a specific value of
$\eta$, the choice between the stable steady states is history-dependent,
i.e., depends on initial conditions \cite{key-22}. If the value of
$\eta$ is initially low, the system ends up in the low $u$ state.
As $\eta$ increases, the system enters the region of bistability
but continues to be in the low expression state till a bifurcation
point is reached. At this point, a discontinuous jump to the high
$u$ state occurs and the system becomes monostable. Bistability is
accompanied by hysteresis , i.e., the value of $\eta$ at which the
switch from the low to the high expression state occurs is greater
than the value of $\eta$ ( the lower bifurcation point ) at which
the reverse transition takes place. The two stable branches are separated
by a branch of unstable steady states (dash-dotted line) which are
not experimentally accessible. There are now several known systems
in which bistability and hysteresis have been observed experimentally
\cite{key-9,key-13,key-23,key-24,key-25,key-26,key-27}. Figure 3(b)
shows the plot of $u$ versus $\eta$ for the same parameter values
as in figure 3(a) except that the copy number of the $GS$ gene is
reduced from two to one, i.e., $n_{S}$ has the value $1$. A comparison
of figures 3(a) and (b) shows that with reduced copy number the extent
of the region of bistability in considerably increased. The same conclusion
is reached when the steady state values of $u$ are plotted versus
the parameter $\beta.$ The region of bistability is lower in extent
when the parameter $\mu,$ a measure of the autorepression strength,
is increased from zero. The value of $\mu$ is changed by modifying
the value of $K_{R}$ (equation (10)), the binding constant for repressor
binding at the $GS$ gene. Figure 4 shows the phase portrait corresponding
to equations (8) and (9) with the parameter values $\xi=30,$ $\eta=30,$
$\alpha=0.002857,$ $\beta=0.002857$ and $\mu=0.$ The system is
bistable for the parameter values quoted. The nullclines, obtained
by putting $\frac{du}{d\tau}\,=\,0,$ $\frac{dv}{d\tau}\,=\,0,$ intersect
at three points, the fixed points of the dynamics. The lower and upper
fixed points are stable whereas the intermediate fixed point is unstable,
in fact, a saddle node \cite{key-36}. The stable manifold of the
saddle node divides the $uv-$phase space into two basins of attraction.
Trajectories starting in the lower (upper) basin of attraction end
up at the lower (upper) stable fixed point as shown in figure 4. A
trajectory initiated on the stable manifold stays on it and ends at
the saddle node. A typical trajectory asymptotically approaches the
unstable manifold as $t\,\rightarrow\,\infty.$ A trajectory is obtained
by plotting the values of $u$ and $v$ at different time points,
determined by solving equations (8) and (9). The arrow direction on
a trajectory denotes increasing time.

\begin{table}
\begin{tabular}{>{\raggedright}p{2.5in}>{\raggedright}p{3in}}
\textbf{Parameter/Rescaled Parameter}&
\textbf{Meaning/Defining Formula}\tabularnewline
$J_{P0},\, J_{S0}$&
Rate constants for basal protein synthesis \tabularnewline
$J_{P},\, J_{S}$&
Rate constants for activated protein synthesis\tabularnewline
$K_{P},\, K_{S}$&
Binding constants of protein dimers $S2$ and $P2$\tabularnewline
$K_{PP},\, K_{SS}$&
Binding constants for the binding of protein dimers $S2$ and $P2$
at the genes $GP$and $GS$\tabularnewline
$\gamma_{P},\,\gamma_{S}$&
Rate constants for protein degradation\tabularnewline
$K_{R}$&
Binding constant for repressor dimer binding at gene $GS;$ $K_{R}$
thus denotes the strength of repression \tabularnewline
$\eta$&
$\eta=\frac{J_{S}}{J_{S0}}$, ratio of activated and basal rate constants
for synthesis of S proteins \tabularnewline
$\xi$&
$\xi=\frac{J_{P}}{J_{P0}},$ ratio of activated and basal rate constants
for synthesis of $P$ proteins \tabularnewline
$\mu$&
$\mu=(\frac{J_{S0}}{\gamma_{S}})^{2}\, K_{S}K_{R};$ with $J_{S0},\,\gamma_{S}$
and $K_{S}$ kept fixed, $\mu$ can be varied by changing $K_{R}$
thus providing a measure of repression strength \tabularnewline
$\alpha$&
$\alpha=(\frac{J_{S0}}{\gamma_{S}})^{2}\, K_{PP}K_{S}$ \tabularnewline
$\beta$&
$\beta=(\frac{\mbox{J}_{P0}}{\gamma_{P}})^{2}\, K_{SS}K_{P}$ \tabularnewline
$T,\, M,\, N$&
abbreviations defined in equations (3) and (4)\tabularnewline
\end{tabular}

\caption{Parameters, rescaled parameters, their meanings and defining formulae }
\end{table}

Figures 5(a) shows the plot of $\xi$ versus $\eta$ exhibiting regions
of monostability and bistability. The parameter values are the same
as before with $\alpha=\beta=0.002857$ and $\mu=0$. The regions
of bistability, enclosed within the red and black curves, correspond
to $n_{S}=1$ and $n_{S}=2$ respectively. The difference in the locations
of the two loops in the logarithmic plots clearly shows that the bistable
region is of greater extent when the gene copy number is reduced from
two to one. The region of bistability is decreased in extent when
autorepression is taken into account (Figure 5(b) with $\mu=0.005).$
Figure 6 shows the $\mu-\beta$ plot with the regions of bistability
falling within the red ($n_{S}=1$) and black ($n_{S}=2$) curves
respectively. The value of $\mu$ is changed by varying $K_{R}$ (equation
10) with $\mu=0.08\; K_{R}$.

\begin{figure}
\begin{centering}
\includegraphics[bb = 0 0 200 100, draft, type=eps]{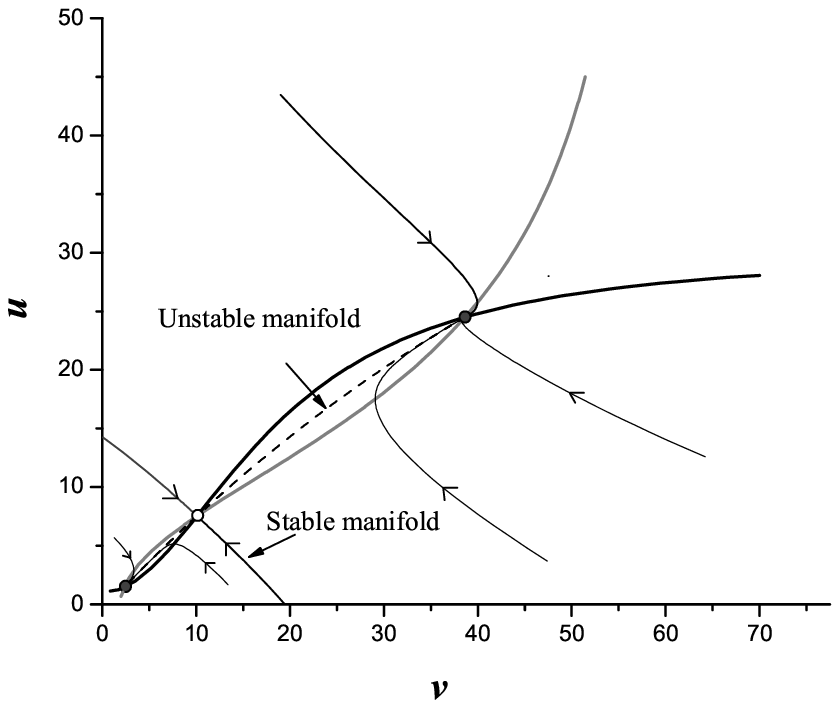}
\par\end{centering}

\caption{Phase portrait described by equations (8) and (9). The dark and light
solid lines represent the nullclines intersecting at three fixed points.
The stable and unstable fixed points are denoted by solid and empty
circles respectively. The stable manifold divides the phase space
into two basins of attraction. Some typical trajectories are shown
with arrow directions denoting increasing time.}
\end{figure}

A major advantage of combining a double positive feedback loop operating
between two genes with autorepression of the expression of one of
the genes lies in dosage compensation \cite{key-16}. This relates
to the fact that the fall in steady state protein levels, brought
about by a reduction in the gene copy number, is less when autorepression
is included, compared to the case when there is no autorepression.
A measure of dosage compensation is provided by the quantity $G$,
termed percentage gain, defined as \begin{equation}
G(\mu)=\frac{x_{1}(\mu)-x_{1}(\mu=0)}{x_{1}(\mu=0)}\times100\end{equation}

\begin{figure}
\includegraphics[clip,scale=0.8,bb = 0 0 200 100, draft, type=eps]{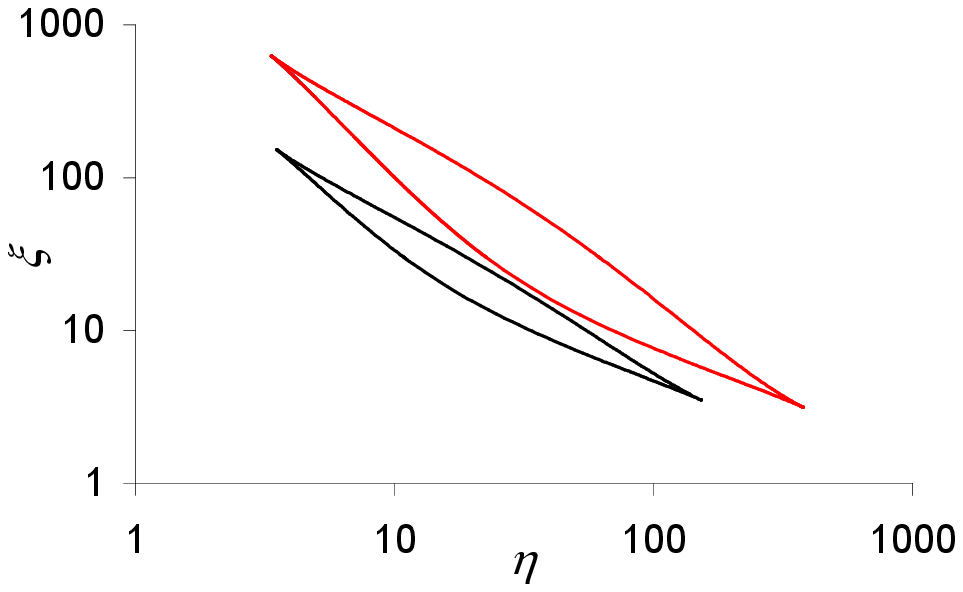}\includegraphics[clip,scale=0.8,bb = 0 0 200 100, draft, type=eps]{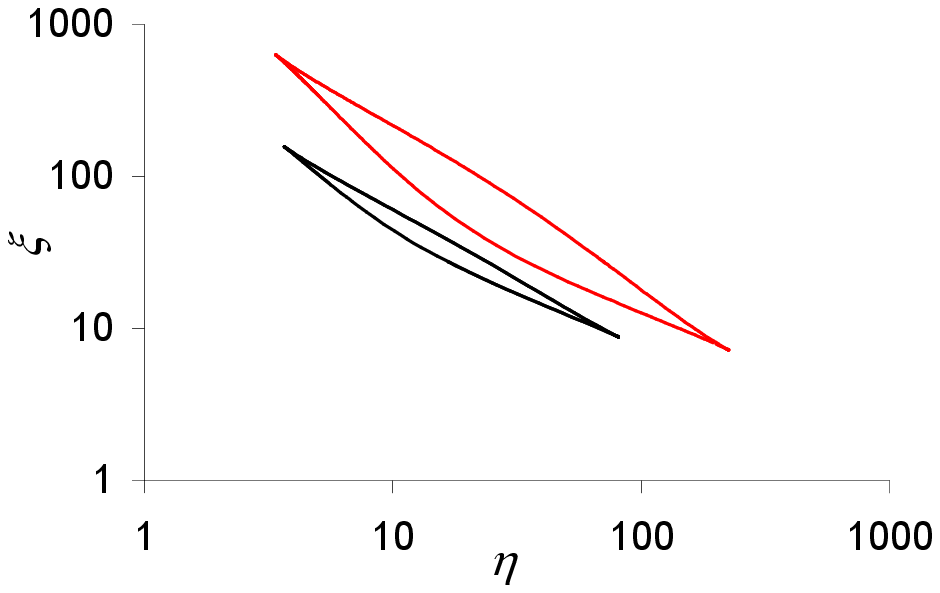}

\caption{Plots of $\xi$ versus $\eta$ showing regions of monostability and
bistability when the parameter $\mu,$ a measure of the autorepression
strength, is zero (a) and 0.005 (b). The regions of bistability are
enclosed within the red and black curves with gene copy numbers $n_{P}=2$,
$n_{S}=1$ and $n_{P}=2$, $n_{S}=2$ respectively. }
\end{figure}

where $x_{1}$ denotes the steady state concentration of $S$ proteins
when the copy number of the $GS$ gene, $n_{S}$, is one. The parameter
$\mu$ is a measure of the repression strength. $G$ is calculated
by keeping the mean level of $S$ proteins to be the same in the two
cases $\mu=0$ and $\mu\neq0$ when $n_{S}=2$. This is achieved by
adjusting the binding constant $K_{PP}$ contained in the parameter
$\alpha$ in equation (10). The other parameter values are $\eta=\xi=30.0$,
$J_{S0}=J_{P0}=2.0$, $\gamma_{S}=\gamma_{P}=1.0$ and $\beta=0.002857$.
Figure 7 shows the plot of $G$ versus $\mu$ for the parameter values
mentioned. As $\mu$ increases from zero, there is initially a sharp
increase in the value of G followed by a slower growth which ultimately
leads to a near-saturation of G values. The results obtained in the
deterministic approach provide insight on the advantages of autorepression
in the non-occurrence of the genetic disorder MODY. The normal functioning
of pancreatic $\beta$-cells requires the $HNF-1\alpha$ and $HNF-4\alpha$
protein levels to be high, i.e., the two-gene system should be in
the ON state. The genesis of MODY lies in a substantial fraction of
the $\beta$-cells being in the OFF state. This is brought about by
mutations in the $hnf-1\alpha$ and $hnf-4\alpha$ genes giving rise
to a fall in the steady state protein levels. In terms of the two-gene
model studied by us, the monostable high state, in which the levels
of the P and S proteins are both high, represents the ON state of
normal $\beta$-cells. The system may enter a region of bistability,
in which both the ON and OFF states are possible, due to the loss
of a gene copy brought about by mutations. We will show in the next
section that fluctuations in the protein levels are responsible for
transitions between the ON and OFF states. In the deterministic scenario,
the major advantages of the autorepressive feedback loop appear to
be dosage compensation (figure 7) as well as a lesser possibility
of the system being in the bistable region due to a reduction in gene
copy number. The continuance of the system in the monostable high
state ensures the normal functioning of cells. Similar conclusions
are reached if the gene copy number $n_{P}$ is reduced from two to
one. There is, however, an asymmetry in the $S$ and $P$ protein
levels as the expression of the gene $GP$ is not autorepressed.

\begin{figure}
\begin{centering}
\includegraphics[clip,scale=0.5,bb = 0 0 200 100, draft, type=eps]{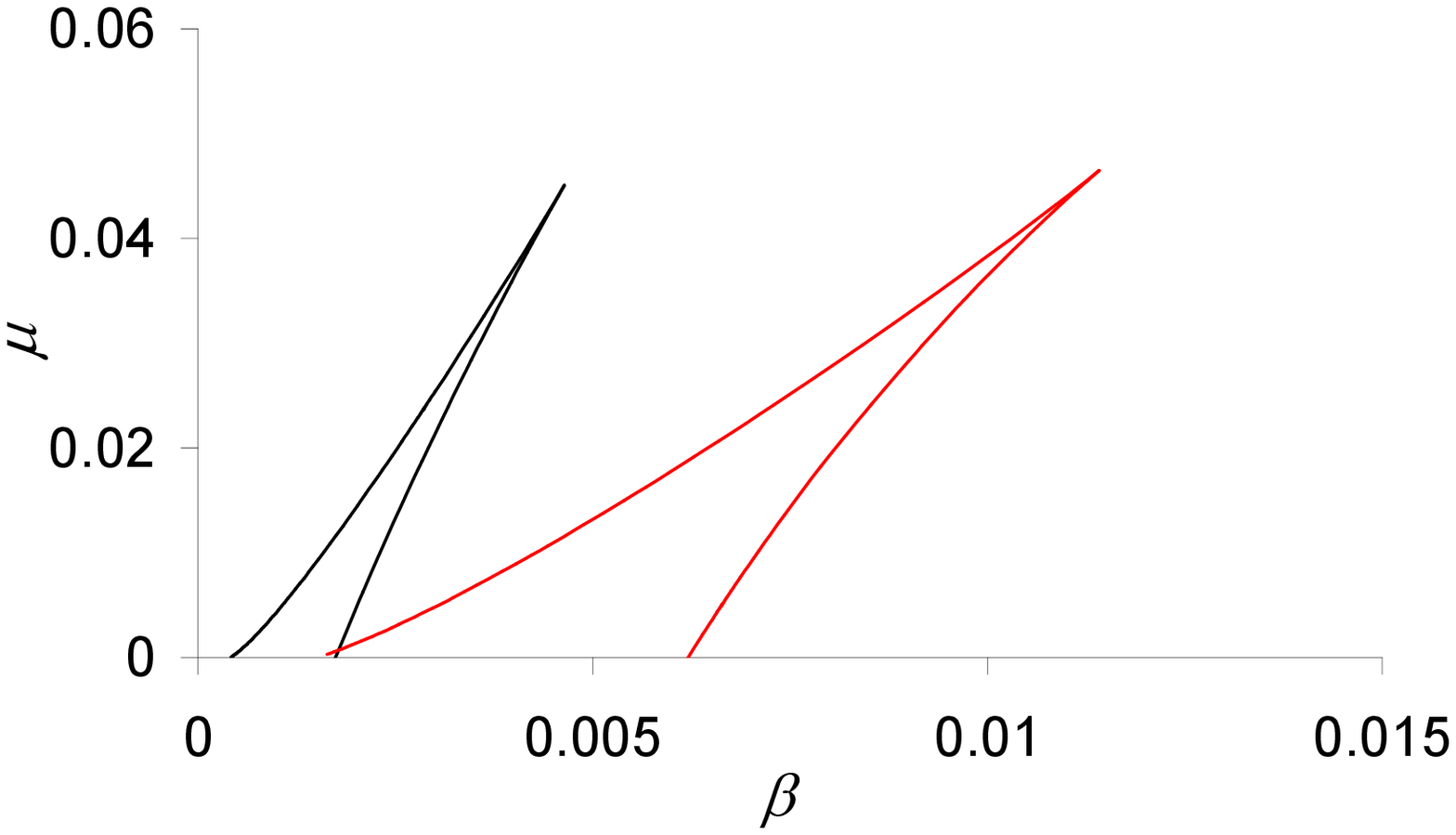}
\par\end{centering}

\caption{$\mu-\beta$ phase diagram with the regions of bistability falling
within the red ($n_{P}=2$, $n_{S}=1$ ) and black ($n_{P}=2$, $n_{S}=2$)
curves respectively. }
\end{figure}

\begin{figure}
\begin{centering}
\includegraphics[bb = 0 0 200 100, draft, type=eps]{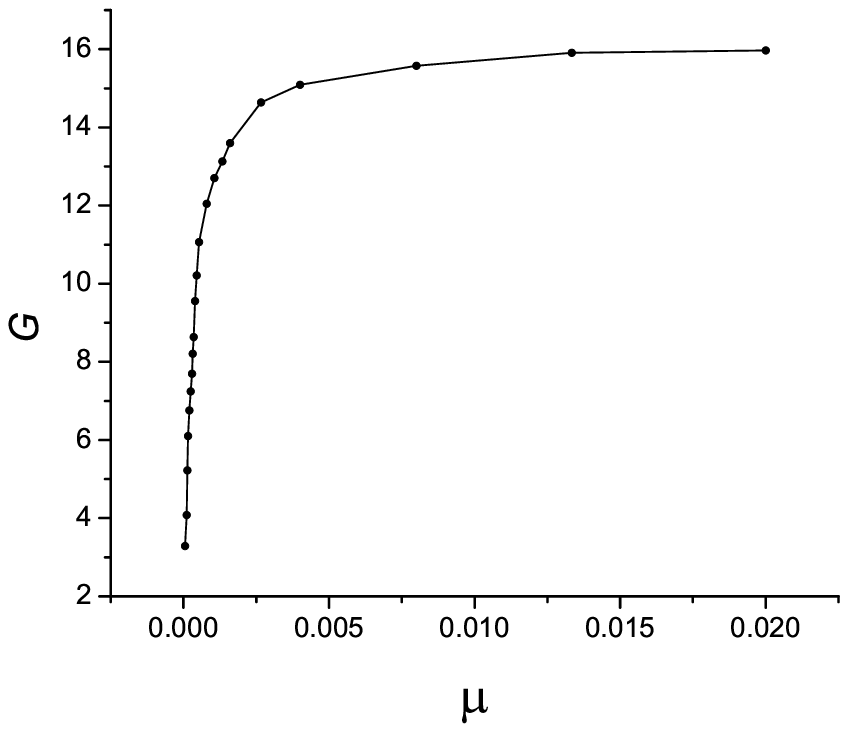}
\par\end{centering}

\caption{Plot of percentage gain $G$ (equation 11) versus $\mu$, a measure
of the autorepression strength.}
\end{figure}

\section{Stochastic Approach}

Consider the two-gene network to be originally in the monostable high
state. In the deterministic formalism, the system continues to be
in the high, i.e., ON state even if it enters the region of bistability
due to the loss of a gene copy. This is due to history dependence,
since the system is initially in the ON state it continues to be in
the ON state in the bistable region. The protein levels corresponding
to the ON state are, however, lower in magnitude in the bistable region.
In the pancreatic $\beta$-cells, the occurrence of MODY is possible
only when a sizable fraction of cells is in the OFF state. The ON$\rightarrow$OFF
and OFF$\rightarrow$ON transitions can be understood only when stochasticity
in gene expression is taken into account. We now give a simple physical
picture of the origin of stochastic transitions \cite{key-6}. In
the case under consideration, the dynamical variables are the protein
concentrations $u$ and $v$. In the case of deterministic time evolution,
trajectories starting in individual basins of attraction stay confined
to the specific basins with no possibility of a trajectory crossing
from one basin to another. In the stochastic approach, the trajectories
are no longer deterministic as the dynamical variables $u(t)$ and
$v(t)$ are fluctuating. In the deterministic case, given the initial
state defined by $(u(t=0),v(t=0))$, the trajectory is fixed in the
$uv$-phase space. In the stochastic case, different trajectories
are generated in repeated trials. A transient fluctuation, if sufficiently
strong, switches the system dynamics from one basin to the other brought
about by the excursion of the trajectory across the boundary separating
the two basins of attraction. In terms of the pancreatic $\beta$-cells,
the switch to the OFF state hampers the normal functioning of the
cells.

\begin{figure}
\begin{centering}
\includegraphics[clip,bb = 0 0 200 100, draft, type=eps]{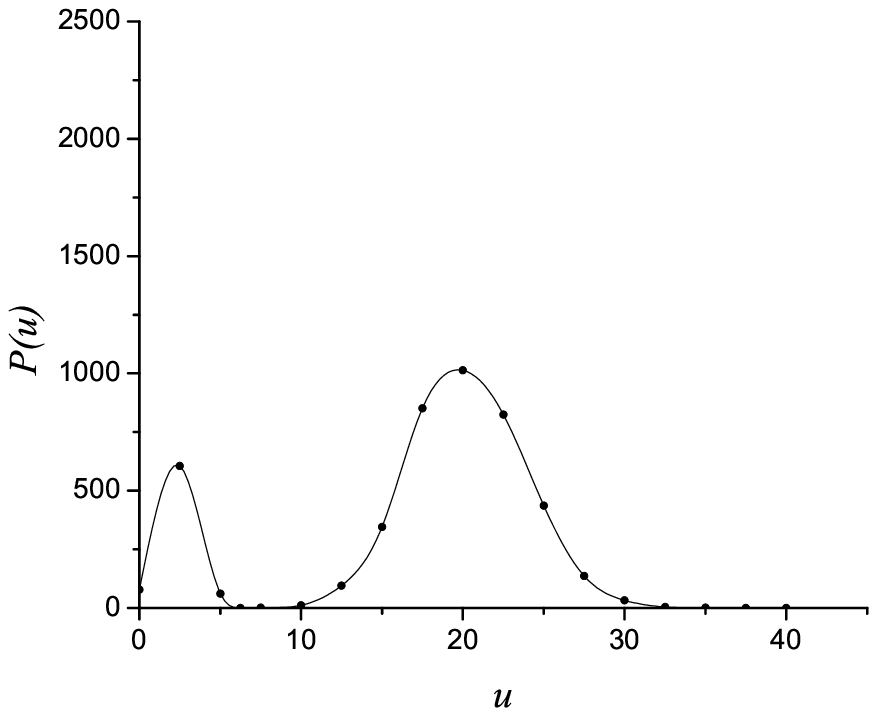}\includegraphics[clip,bb = 0 0 200 100, draft, type=eps]{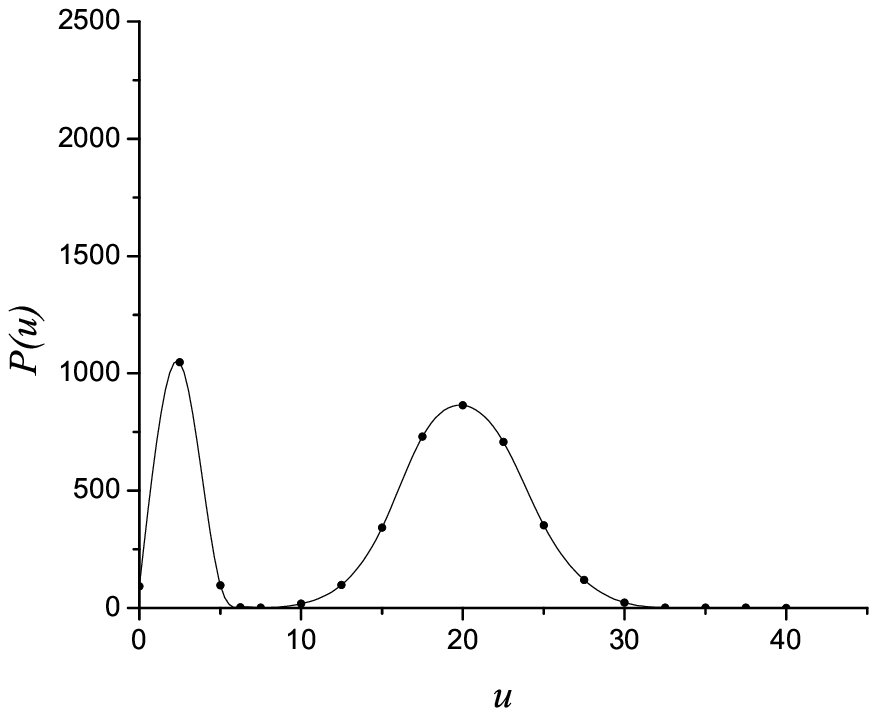}
\par\end{centering}

\begin{centering}
\includegraphics[clip,bb = 0 0 200 100, draft, type=eps]{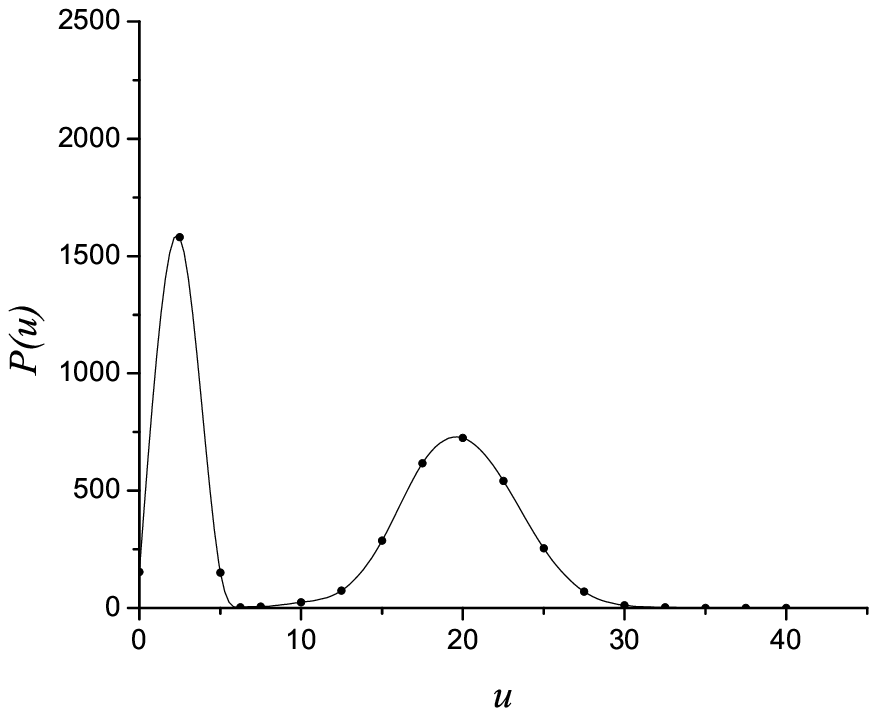}\includegraphics[clip,bb = 0 0 200 100, draft, type=eps]{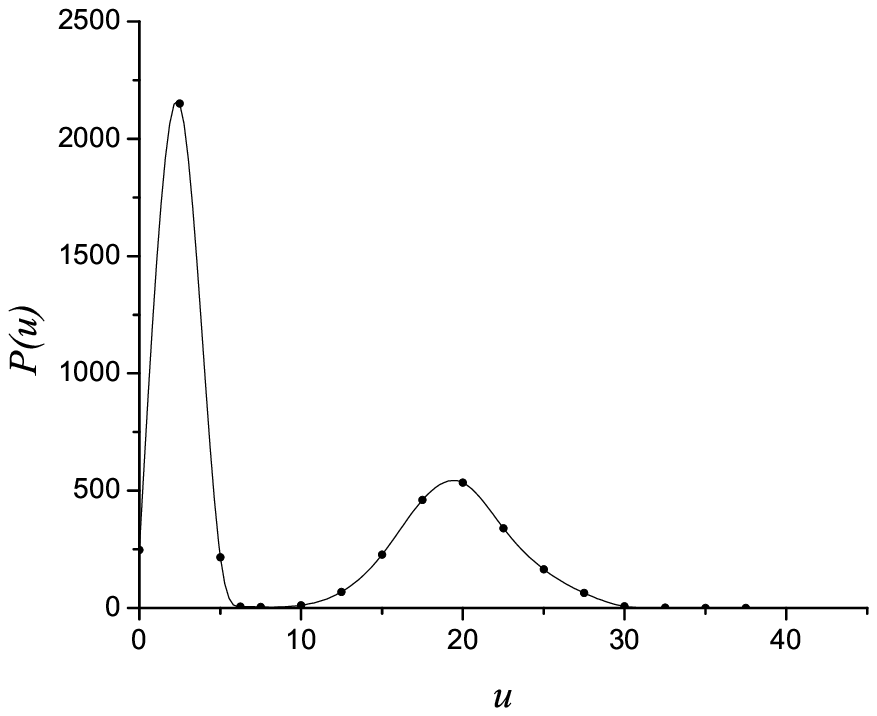}
\par\end{centering}

\caption{Distribution of steady state $GS$ protein levels, $P(u)$, in an
ensemble of $4500$ cells for repressor strengths (a) $\mu=0.000027$,
(b) $\mu=0.00002$, (c) $\mu=0.00001$ and (d) $\mu=0.0$ respectively.}
\end{figure}

For proper regulatory functions as transcription factors, the $HNF-1\alpha$
and $HNF-4\alpha$ protein levels are high with an optimal value as
excessive protein amounts are known to be harmful rather than beneficial
\cite{key-16}. In this context, it is pertinent to undertake a comparison
of the functional characteristics of two-gene network models with
and without the autorepressive loop and with the mean protein levels
kept at the same high values in the two cases. The last condition
ensures the normal functioning of the cells in both the cases. In
section 2, we have identified certain advantages of the autorepressive
loop as regards the system dynamics in a deterministic framework.
Our goal is now to identify the desirable features of the model incorporating
both a double positive feedback loop and an autorepressive loop taking
the stochastic aspects of the dynamics into consideration. This is
done with the help of a detailed computer simulation based on the
Gillespie algorithm \cite{key-28}. The algorithm enables one to keep
track of the stochastic time evolution of the system. The different
biochemical reactions considered in the simulation are depicted in
figures 2(a) and 2(b). The reactions are sixteen in number and are
given by \begin{equation}
GS+P2\rightarrow GS^{*}\end{equation}
\begin{equation}
GS^{*}\rightarrow S\end{equation}
\begin{equation}
GS\rightarrow S\end{equation}
\begin{equation}
S+S\rightarrow S2\end{equation}
\begin{equation}
GS^{*}\rightarrow GS+P2\end{equation}
\begin{equation}
S2\rightarrow S+S\end{equation}
\begin{equation}
S\rightarrow\Phi\end{equation}
\begin{equation}
GP+S2\rightarrow GP^{*}\end{equation}
\begin{equation}
GP^{*}\rightarrow P\end{equation}
\begin{equation}
GP\rightarrow P\end{equation}
\begin{equation}
P+P\rightarrow P2\end{equation}
\begin{equation}
GP^{*}\rightarrow GP+S2\end{equation}
\begin{equation}
P2\rightarrow P+P\end{equation}
\begin{equation}
P\rightarrow\Phi\end{equation}
\begin{equation}
GS+S2\rightarrow GSS2\end{equation}
\begin{equation}
GSS2\rightarrow GS+S2\end{equation}

The different symbols are as explained in section 2. The stochastic
rate constants, associated with the reactions, are C(i), i=1,...,16,
in the appropriate units. The results of the simulation are shown
in figures 8-9. Figures 8(a)-(d) show the distribution of $GS$ protein
levels, $P(u)$, in an ensemble of 4500 cells for repressor strengths
$\mu=0.000027,\;0.00002,\;0.00001$ and $0.0$ respectively after
a simulation time of $tmax=2000$ time units. The gene copy numbers
are $n_{P}=2$ and $n_{S}=1$ so that the system is in the region
of bistability. The values of the stochastic rate constants are $C(2)=56.0,C(3)=2.0,C(4)=4.0,$
$C(5)=280.0,\: C(6)=100.0,\: C(7)=1.0,\: C(8)=10.0,\:$$C(9)=50.0$,$\: C(10)=2.0$,$\: C(11)=4.0$,$\: C(12)=280.0$,
$C(13)=100.0,\:$$C(14)=1.0,\: C(15)=10.0$. The value of $\mu$ is
changed by varying the stochastic rate constant $C(16)$. The value
of the rate constant $C(1)$ is changed to keep the mean protein levels
to be the same when $n_{S}=2,\: n_{P}=2$ for all values of $\mu$.
The value $\mu=0$ implies that only the double positive feedback
loop contributes to the dynamics. The distribution $P(u)$ is found
to be bimodal, i.e., has two distinct peaks corresponding to the OFF
and ON states. In all the cases, the cells are in the ON state at
time $t=0.$ One finds that the fraction of cells in the OFF state
decreases as the value of $\mu$ increases. In fact, when $\mu=0,$
the number of cells which are in the OFF state is larger than that
in the ON state. Since initially all the cells are in the ON state,
a large number of ON $\rightarrow$ OFF state transitions occur during
the simulation time. For $\mu=0,$ the reverse transition is, however,
much rarer. The role of the autorepressive loop thus appears to be
to reduce the number of stochastic transitions from the ON to the
OFF state. This makes the occurrence of MODY, brought about by a sizeable
fraction of the cell population existing in the OFF state, less probable.
There are two distinct time scales over which protein fluctuations
occur. The probability distribution $P(u)$ versus $u$ has a two-peaked
structure. Fluctuations on a short time scale confine the $u$ values
to lie predominantly within individual peaks. The long time scale
corresponds to the time at which large fluctuations occur bringing
about transitions between states belonging to different peaks. The
{}``escape time'' is often very large and a quantitative measure
is provided by the mean first passage time $\tau$ \cite{key-37}.
In the present case, the values of $\tau_{ON\rightarrow OFF}$ and
$\tau_{OFF\rightarrow ON}$ are quite large for different values of
$\mu.$ The maximum simulation time $tmax$ is $2000$ time units
for all values of $\mu.$ For $\mu=0,$ $\tau_{ON\rightarrow OFF}$
is around $1000$ time units whereas $\tau_{OFF\rightarrow ON}$ is
even larger. As $\mu$ increases, $\tau_{ON\rightarrow OFF}$ increases
whereas $\tau_{OFF\rightarrow ON}$ decreases. For $\mu=0.0005,$
$\tau_{ON\rightarrow OFF}$ is as large as $10^{7}s.$ Because of
large escape times, the probability distribution $P(u)$ versus $u$
is metastable on a large time scale \cite{key-37}. Over shorter periods
of time, the shape of the distribution remains more or less invariant. 

The plots in figure 9 are obtained for an ensemble of 4500 cells.
For gene copy numbers $n_{P}=2$ and $n_{S}=2$, the mean protein
levels are adjusted to be the same irrespective of the values of $\mu$
. The parameter values are so chosen that the system is in the monostable
high region. On reduction of $n_{S}$ to $1$ (one copy of the GS
gene), the system enters the region of bistability and is in the ON
state at time $t=0$. After a period $T=2000$ time units of stochastic
time evolution, the percentage of cells in the OFF state is determined.
The red curve shows this percentage as a function of the repression
strength $\mu$. The drop in the percentage of cells in the OFF state
is found to be exponential. The black curve shows the percentage of
cells in the ON state after $T=2000$ time units, with all the cells
being initially in the OFF state. One finds that with increasing $\mu$,
the fraction of cells in the ON state becomes larger. The autorepressive
loop has the effect of making the ON state more stable and the OFF
state more unstable. This feature enhances the probability of the
nonoccurrence of MODY as there are infrequent transitions from the
ON to the OFF state. On the other hand, the system has a lesser probability
of remaining stuck in the OFF state compared to the case when there
is no autorepressive loop.

\begin{figure}
\begin{centering}
\includegraphics[bb = 0 0 200 100, draft, type=eps]{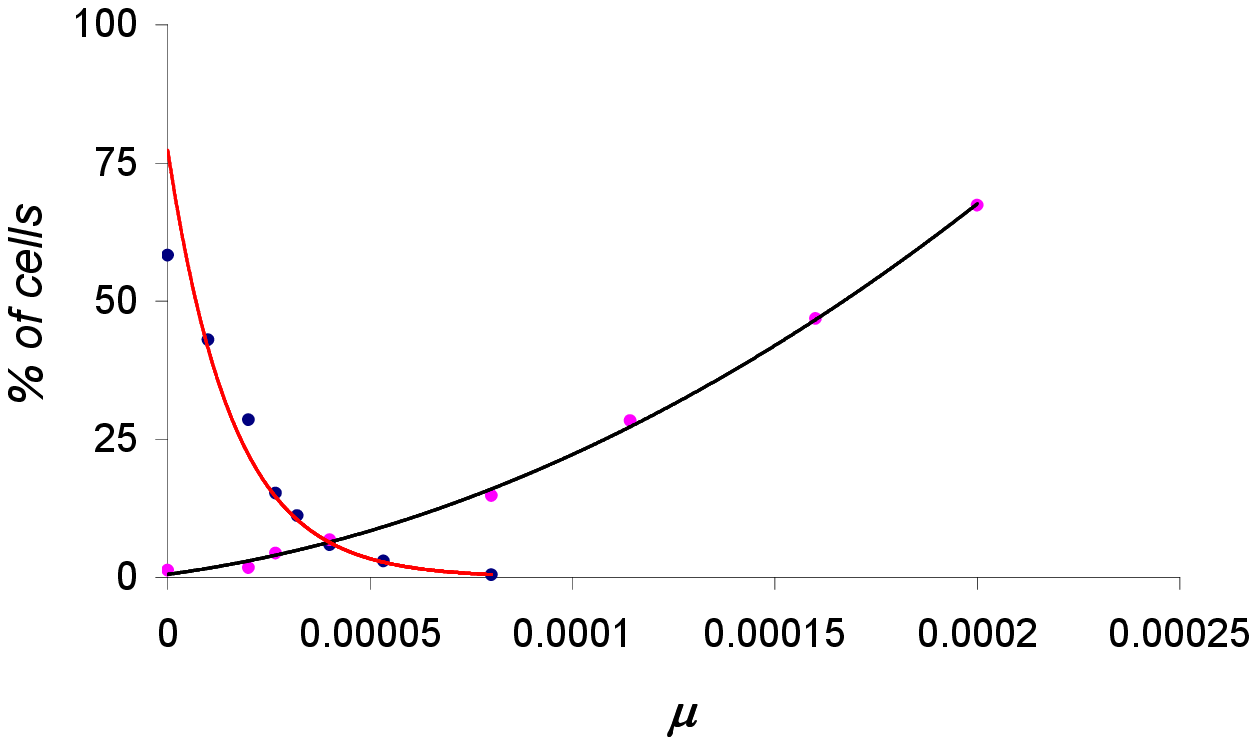}
\par\end{centering}

\caption{For gene copy numbers $n_{P}=2,\: n_{S}=1$ and after a time interval
$T=2000$ time units of stochastic time evolution, the percentage
of cells in an ensemble of $4500$ cells in the OFF state (red curve)
versus the repression strength $\mu$ with all the cells being in
the ON state at time $t=0$. The black curve shows the percentage
of cells in the ON state versus $\mu$ with all the cells being in
the OFF state at $t=0$.}
\end{figure}

\section{Conclusion and Outlook}

In this paper, we have studied the functional characteristics of a
motif consisting of a double positive feedback loop operating between
two genes and a negative feedback loop in which the protein product
of one gene represses its own synthesis. The motif appears in the
gene regulatory network controlling the pancreatic $\beta$-cell function
\cite{key-16}. Loss of a gene copy due to mutations has been shown
\cite{key-16} to be responsible for abnormal $\beta$-cell function
resulting in MODY. We have studied the effect of reduced gene copy
number on the dynamics of the model describing the two-gene motif.
In a deterministic formalism based on differential rate equations,
we identified regions of bistability in appropriate parameter regions.
The stable steady states, designated as the OFF and ON states, correspond
respectively to low and high protein levels. The normal $\beta$-cells
are expected to be in the monostable ON state. The occurrence of MODY
is brought about by a fraction of $\beta$-cells being in the OFF
state. The ON$\rightarrow$OFF switch can occur only in the bistable
region. Negative feedback reduces the extent of the bistable region
making it less likely that the cellular state falls in this part of
the phase diagram. The region of bistability, however, increases in
size on reduction of the gene copy number making the ON$\rightarrow$OFF
transitions more probable. Negative feedback also produces a mechanism
of dosage compensation (figure 7). The results hold true for a wide
range of parameter values. Since switching to the OFF state is detrimental,
one would have thought, from an evolutionary point of view, that two
genes which are constitutively ON would be more appropriate. In reality,
the genes $hnf-1\alpha$ and $hnf-4\alpha$ form a positive feedback
loop. Cross-regulation between the two genes is established when the
pancreatic cells receive signals to differentiate \cite{key-16}.
The positive feedback loop provides a stable mechanism of gene expression
since the two genes reset each other's activity to the functional
state under physiological perturbations. This serves to self-perpetuate
the activity of the two genes and their targets in the pancreatic
$\beta-$cells. Normal functioning of these cells requires both the
protein levels to be high. The system of two genes that are constitutively
ON are less robust under physiological perturbations since there is
no resetting mechanism by which both the genes are in the functional
ON state. The theoretical suggestions of bistability due to the existence
of a positive feedback loop \cite{key-16,key-35}, backed up by the
results of our mathematical model, should be tested in actual experiments. 

The ON$\rightarrow$OFF switch is brought about by protein fluctuations
the origin of which lies in stochastic gene expression. Our major
finding is that negative feedback makes the ON$\rightarrow$OFF transitions
less probable and the OFF$\rightarrow$ON transitions more probable.
Thus the function of the negative feedback appears to be to protect
the normal $\beta$-cell function since the cell is more likely to
be in the ON state in this case. The asymmetric response to fluctuations
prevents switching off and facilitates switching on of the high expression
state. In the deterministic scenario, one finds that the difference
between the ON state and the unstable steady state protein level increases
as the autorepression strength is increased whereas the difference
between the unstable steady state and OFF state protein levels decreases
on increasing the autorepression strength. This may explain the asymmetry
in the ON$\rightarrow$OFF and OFF$\rightarrow$ON switches when stochasticity
is taken into account. For moderate strengths of autorepression, the
system is locked in the ON state for extremely long times. In our
simulations, we did not encounter ON$\rightarrow$OFF switches for
very long trajectories ($\sim10^{7}$ seconds) with $\mu=0.0005$.
This translates into lifetimes measured in years and explains the
delayed onset of the diabetic phenotype \cite{key-16}. The phenotype
generally appears after several years indicating that the activation
of the ON$\rightarrow$OFF switch is rare. The average age at which
MODY is manifest could thus be dictated by the probability that a
sufficient number of $\beta$-cells is locked in the OFF state. We
have considered the simplest form of negative autoregulation in our
two-gene model. There are recent suggestion that negative autoregulation
of the HNF-4$\alpha$ gene in the pancreatic $\beta$-cells may be
more complex \cite{key-29}. Also, the number of transcription factor
binding sites of the two genes is not known with certainty. Cooperative
binding at multiple sites is expected to promote the stability of
the gene expression states. Our two-gene motif constitutes a minimal
model which seeks to explains the desirable features of combining
a double positive feedback loop with an autorepressive loop vis-\'a-vis
the normal functional activity of $\beta$-cells. The insight gained
from the model study is expected to provide a basis for the investigation
of more complex cases.

\newpage

Figure Captions

Fig1. The two-gene network model. The protein products of the $GP$
and $GS$ genes activate each other's synthesis. There is also an
autorepressive loop in which the proteins of the $GS$ gene repress
their own synthesis.

Fig2. The reaction kinetic scheme of the two-gene model. The meanings
of the symbols are explained in the text.

Fig3. $u$ versus $\eta$ curves showing bistability and hysteresis.
The solid (dashed) lines represent stable (unstable) steady states
for gene copy numbers (a) $n_{P}=2$, $n_{S}=2$ and (b) $n_{P}=2$,
$n_{S}=1$ . The parameter $\mu$, a measure of the autorepression
strength, is zero.

Fig4. Phase portrait described by equations (8) and (9). The dark
and light solid lines represent the nullclines intersecting at three
fixed points. The stable and unstable fixed points are denoted by
solid and empty circles respectively. The stable manifold divides
the phase space into two basins of attraction. Some typical trajectories
are shown with arrow directions denoting increasing time.

Fig5. Plots of $\xi$ versus $\eta$ showing regions of monostability
and bistability when the parameter $\mu,$ a measure of the autorepression
strength, is zero (a) and 0.005 (b). The regions of bistability are
enclosed within the red and black curves with gene copy numbers $n_{P}=2$,
$n_{S}=1$ and $n_{P}=2$, $n_{S}=2$ respectively. 

Fig6. $\mu-\beta$ phase diagram with the regions of bistability falling
within the red ($n_{P}=2$, $n_{S}=1$ ) and black ($n_{P}=2$, $n_{S}=2$)
curves respectively. 

Fig7. Plot of percentage gain $G$ (equation 11) versus $\mu$, a
measure of the autorepression strength.

Fig8. Distribution of steady state $GS$ protein levels, $P(u)$,
in an ensemble of $4500$ cells for repressor strengths (a) $\mu=0.000027$,
(b) $\mu=0.00002$, (c) $\mu=0.00001$ and (d) $\mu=0.0$ respectively.

Fig9. For gene copy numbers $n_{P}=2,\: n_{S}=1$ and after a time
interval $T=2000$ time units of stochastic time evolution, the percentage
of cells in an ensemble of $4500$ cells in the OFF state (red curve)
versus the repression strength $\mu$ with all the cells being in
the ON state at time $t=0$. The black curve shows the percentage
of cells in the ON state versus $\mu$ with all the cells being in
the OFF state at $t=0$.

\newpage

\begin{thebibliography}{10}
\bibitem{key-1}Alon U 2007 Network motifs: theory and experimental
approaches Nat. Rev Genet. \textbf{8} 450-61

\bibitem{key-2}Mitrophanov A Y and Groisman E A 2008 Positive feedback
in cellular control systems Bioessays \textbf{30} 452-55

\bibitem{key-3}Becskei A and Serrano L 2000 Engineering stability
in gene networks by autoregulation Nature \textbf{405} 590-93

\bibitem{key-4}Becskei A, Seraphin B and Serrano L 2001 Positive
feedback in eukaryotic gene networks: cell differentiation by graded
to binary response conversion EMBO J. \textbf{20} 2528-35

\bibitem{key-5}Maeda Y T and Sano M 2006 Regulatory dynamics of synthetic
gene networks with positive feedback J. Mol. Biol. \textbf{359} 1107-24

\bibitem{key-6}Karmakar R and Bose I 2007 Positive feedback, stochasticity
and genetic competence Phys. Biol. \textbf{4} 29-37

\bibitem{key-7}Hornung G and Barkai N 2008 Noise propagation and
signaling sensitivity in biological networks: a role for positive
feedback PLOS Coputatational Biology \textbf{4} 0055-61

\bibitem{key-8}Stekel D J and Jenkins D J 2008 Strong negative self
regulation of prokaryotic transcription factors increases the intrinsic
noise of protein expression BMC Systems Biology \textbf{2}

\bibitem{key-9}Gardner T S, Cantor C R and Collins J J 2000 Construction
of a genetic toggle switch in Escherichia coli Nature \textbf{403}
339-42

\bibitem{key-10}Tyson J J, Chen K and Novak B 2001 Network dynamics
and cell physiology Nat. Rev. Mol. Cell Biol. \textbf{2} 908-16

\bibitem{key-11}Brandman O, Ferrell J E, Li R and Meyer T 2005 Interlinked
fast and slow positive feedback loops drive reliable cell decisions
Science \textbf{310} 496-8

\bibitem{key-12}Kim D, Kwon Y K and Cho K H 2007 Coupled positive
and negative feedback circuits form an essential building block of
cellular signalling pathways Bioessays \textbf{29} 85-90

\bibitem{key-13}Acar M, Becskei A and van Oudenaarden A 2005 Enhancement
of cellular memory by reducing stochastic transitions Nature \textbf{435}
228-32

\bibitem{key-14}Kaern M, Elston T C, Blake W J and Collins J J 2005
Stochasticity in gene expression: from theories to phenotypes Nat.
Rev. Genet. \textbf{6} 451-64 

\bibitem{key-15}S\"uel G M, Garcia-Ojalvo J, Liberman L M and Elowitz
M B 2006 An excitable gene regulatory circuit induces transient cellular
differentiation Nature \textbf{440} 545-50

\bibitem{key-16}Ferrer J 2002 A genetic switch in pancreatic $\beta$-cells:
Implications for differentiation and haploinsufficiency \textbf{51}
2355-62

\bibitem{key-34}Kulkarni R N and Kahn C R 2004 HNFs- linking the
liver and pancreatic islets in diabetes Science \textbf{303} 1311-12

\bibitem{key-35}Odom D T et al. 2004 Control of pancreas and liver
gene expression by HNF transcription factors Science \textbf{303}
1378-1381

\bibitem{key-17}Veitia R A 2002 Exploring the etiology of haploinsufficiency
Bioessays \textbf{24} 175-84

\bibitem{key-18}Fodde R and Smits R 2002 A matter of dosage Science
\textbf{298} 761-3

\bibitem{key-19}Cook D L, Gerber A N and Tapscott S J 1998 Modelling
stochastic gene expression: Implications for haploinsufficiency Proc.
Natl. Acad. Sci. \textbf{95} 15641-6

\bibitem{key-20}Bose I and Karmakar R 2005 The Biology of Genetic
Dominance (Georgetown, T X: Landes Bioscience) Chapter 6

\bibitem{key-21}XPPAUT (http://www.math.pitt.edu/\textasciitilde{}bard/xpp/xpp.html)

\bibitem{key-22}Ferrell J E 2002 Self-perpetuating states in signal
transduction: positive feedback, double-negative feedback and bistability
Curr. Opin. Cell Biol. \textbf{14} 140-8

\bibitem{key-23}Sha W, Moore J, Chen K, Lassaletta A D, Yi C S, Tyson
J J and Sible J C 2003 Hysteresis drives cell-cycle transitions in
Xenopus laevis egg extracts Proc. Natl. Acad. Sci. \textbf{100} 975-80 

\bibitem{key-24} Pomerening1 J R, Sontag E D and Ferrell J E 2003
Building a cell cycle oscillator: hysteresis and bistability in the
activation of Cdc2 Nature Cell Biol\textbf{. 5} 346 - 51 

\bibitem{key-25}Ozbudak E M , Thattai M, Lim H N, Shraiman B I, Oudenaarden
van A 2004 Multistability in the lactose utilization network of Escherichia
coli Nature Vol. \textbf{427} 737-40.

\bibitem{key-26} Kramer B P and Fussenegger M 2005 Transgene control
engineering in mammalian cells Methods Mol Biol. \textbf{308} 123-43

\bibitem{key-27}Karmakar R and Bose I 2007 Positive feedback, stochasticity
and genetic competence Phys. Biol. \textbf{4} 29-37 

\bibitem{key-36}Strogatz S H 1994 Nonlinear Dynamics and Chaos: with
Applications to Physics, Biology, Chemistry and Engineering (Cambridge:
Perseus Books)

\bibitem{key-28}Gillespie D T 1977 Exact stochastic simulation of
coupled chemical reactions J. Phys. Chem. 81 2340-61

\bibitem{key-37}Van Kampen N G 1997 Stochastic processes in Physics
and Chemistry 2nd edn (Amsterdam: North-Holland)

\bibitem{key-29}Magenheim J, Hertz R, Berman I, Nousbeck J and Jacob
Bar-Tana J Negative autoregulation of HNF-4$\alpha$ gene expression
by HNF-4$\alpha$1 Biochem. J 384

\end{thebibliography}
\end{document}